\documentclass[vecphys]{svmult}

\usepackage{makeidx}         % allows index generation
\usepackage{graphicx}        % standard LaTeX graphics tool
\usepackage{multicol}        % used for the two-column index
\usepackage[bottom]{footmisc}% places footnotes at page bottom
\makeindex             % used for the subject index
\def\aap{\hbox{A\&A$\;$}}
\def\ApJ{\hbox{ApJ$\;$}}
\def\apj{\hbox{ApJ$\;$}}
\def\nat{\hbox{Nature$\;$}}
\def\spb{\smallskip\par\noindent $\bullet\;$}
\catcode`\@=11
\def\gsim{\ifmmode{\mathrel{\mathpalette\@versim>}}
    \else{$\mathrel{\mathpalette\@versim>}$}\fi}
\def\lsim{\ifmmode{\mathrel{\mathpalette\@versim<}}
    \else{$\mathrel{\mathpalette\@versim<}$}\fi}
\def\@versim#1#2{\lower 2.9truept \vbox{\baselineskip 0pt \lineskip 
    0.5truept \ialign{$\m@th#1\hfil##\hfil$\crcr#2\crcr\sim\crcr}}}
\catcode`\@=12
\def\msun{\hbox{$M_\odot$}}
\def\spb{\smallskip\par\noindent $\bullet\;$}
\begin{document}

\title*{The  Rest-Frame UV Spectrum of Elliptical Galaxies at High Redshift}
\author{Alvio Renzini}
%\and Name of Author\inst{1}}
\institute{INAF--Osservatorio Astronomico di Padova, Padova, Italy}

%\texttt{alvio.renzini@oapd.inaf.it}}
\maketitle

\begin{abstract}
Beyond redshift $\sim 1.4$ the only spectral feature that allows one
to get the redshift of passively evolving galaxies (PEG) with optical
spectrographs is a characteristic structure due to a set of iron and
magnesium lines lines at $\lambda\simeq 2600-2850$ \AA\ in the rest
frame. The same feature permits also to estimate the time elapsed
since the cessation of star formation. Current efforts at observing
high redshift PEGs at the VLT and SUBARU telescopes are briefly reviewed.

\end{abstract}

\section{Introduction}
\label{sec:1}
% Always give a unique label
% and use \ref{<label>} for cross-references
% and \cite{<label>} for bibliographic references
% use \sectionmark{}
% to alter or adjust the section heading in the running head
In the local universe more than half of the stellar mass resides in
passively evolving galaxies (PEG), and above $\sim 10^{11}M_\odot$
PEGs outnumber starforming galaxies by a factor of 10 or more
(e.g. Baldry et al. 2004). As most
PEGs are morphologically classified as ellipticals, and most
ellipticals are PEGs (e.g., Renzini 2006, Scarlata et al. 2007)
sometimes PEGs and ellipticals, or early type galaxies (ETG), are used as
synonyms, though there is no perfect overlap, neither at low nor high
redshift (e.g. Scarlata et al. 2007), This makes PEGs an especially
important class of cosmic objects, and understanding their formation
and evolution is central to the broader issue of galaxy formation in
general.

\begin{figure}
\centering
\includegraphics[width=9cm, angle=-90]{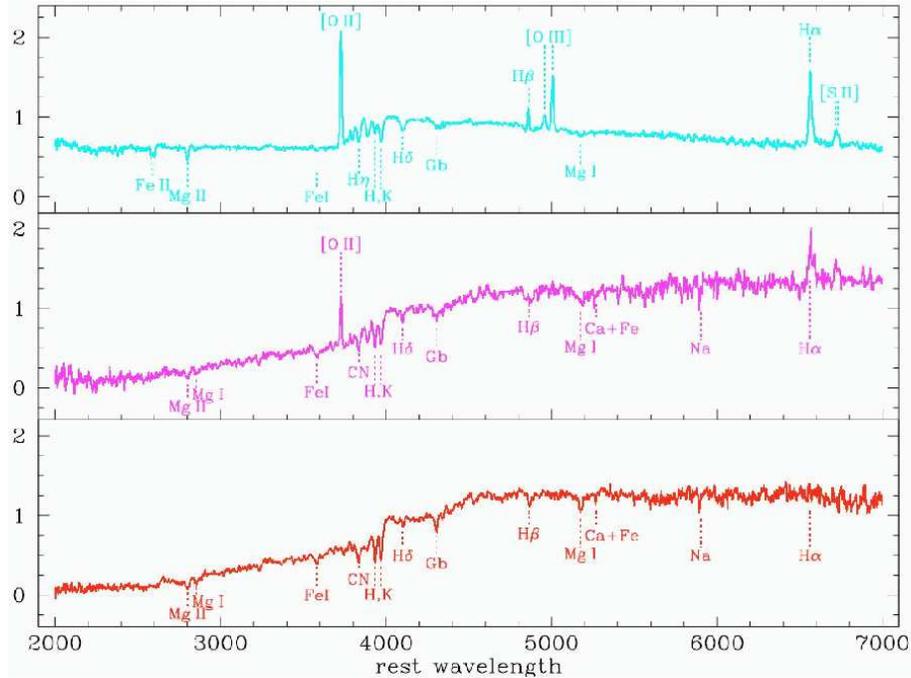}
%\picplace{5cm}{2cm} % Give the correct figure height and width in cm
%
\caption{A set of co-added spectra from the K20 survey (Mignoli et
al. 2005).  The top spectrum refers to actively star-forming galaxies,
the bottom one to purely passive galaxies, and the middle spectrum
refers to mostly passive galaxies with just a sprinkle of star
formation.}
\label{fig:1}       % Give a unique label
\end{figure}

Only in recent years it has become possible to observationally map the
evolution of the PEG population all the way to fairly high redshifts
($z\gsim 1$), as a result of combining space and ground observations
taken at the most powerful facilities. The ultimate aim of these
efforts is that of tracing the evolution with redshift of the number
density and mass function of PEGs, all the way to their disappearance,
i.e. to the epoch when their precursors were still actively star
forming, or were not assembled yet. Thus, to this end three main
issues have to be attacked, namely: 
\spb 
Identify suitable PEG candidates at high redshift, 
\spb
Measure their redshift, and 
\spb
Estimate age, metallicity and mass of their stellar populations.

\smallskip\noindent In this paper I will briefly report the main
results from projects such as the K20 (Cimatti et al.  2002; Mignoli
et al. 2005), GMASS (Cimatti et al., in preparation; Kurk et al., in
preparation), and a joint ESO/SUBARU project (Kong et al. 2006; Daddi
et al., in preparation). Multiobject spectroscopy in the rest-frame
ultraviolet has been so far the main tool that has allowed us to
precisely locate high-$z$ PEGs in space (and time). Quite similar
results are being obtained by the GDDS team, and those are reported by
Pat McCarthy at this meeting.

\section{PEGs beyond $z\sim 1.4$}

The spectroscopic identification of PEGs is relatively easy up to
redshift $\sim 1$. As shown in Figure 1, the CaII H\&K lines together
with the 4000 \AA\ break are the strongest features, and are very
effective for measuring accurate redshifts. Large samples of PEGs have
then been culled up to this redshift (e.g. Zucca et al. 2005; Faber et
al. 2007; Scarlata et al. 2007). However, at higher redshifts these features 
first become contaminated by atmospheric OH emission, and then move beyond
the sensitivity limit of CCD detectors. Hence, other features need to be used
to measure redshifts with optical spectrographs.

\begin{figure}
\centering
\includegraphics[height=8cm]{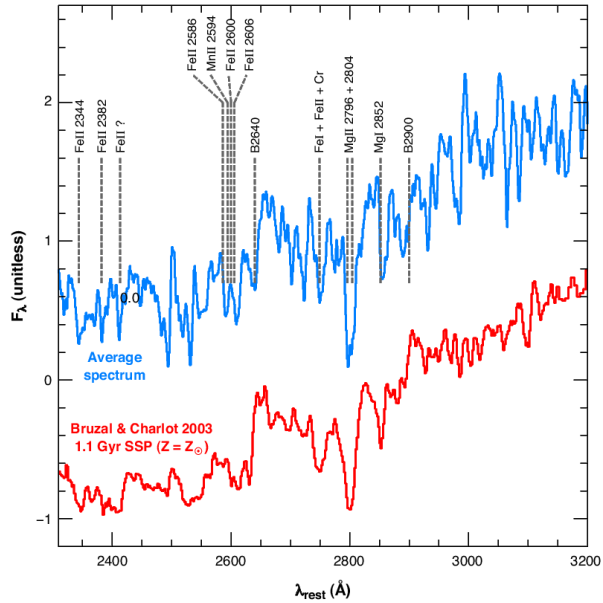}
%\picplace{5cm}{2cm} % Give the correct figure height and width in cm
%
\caption{The coadded rest-frame UV spectrum of 4 PEGs from the K20 survey
(Cimatti et al. 2004). For comparison the synthetic spectrum of a 1.1 Gyr old 
stellar population with solar metallicity is also shown.}
\label{fig:2}       % Give a unique label
\end{figure}

An inspection of the PEG spectrum in Figure 1 shows that to the blue of
the CaII doublet the strongest features are an isolated FeI line at
$\lambda\sim 3580$, and a complex feature extending over the range
$\lambda=2600-2850$, due to various Fe, MgI and MgII lines, that we
call the Mg-UV feature. It is basically thanks to this feature that
all redshifts of PEGs at $z>1.4$ have so far been measured.

The fist such case was for a PEG at $z=1.55$ selected for being a
radiogalaxy (Dunlop et al 1996; Spinrad et al. 1997). Then almost a
decade passed before other PEGs at $z>1.4$ could be spectroscopically
identified (5 such objects by Glazebrook et al.  2004, and 4 objects
by Cimatti et al. 2004). The most exciting aspect of these 2004
discoveries was that the 9 PEGs were found over a combined field of
only $\sim 52$ arcmin$^2$, while the Dunlop et al. object was selected
from a catalog covering virtually the whole sky. The clear implication
was that massive PEGs at high redshift ought to be much more numerous
that many had suspected.

Figure 2 shows the coadded spectrum of the 4 Cimatti et al. objects,
along with a synthetic spectrum of a 1.1 Gyr old simple stellar
population (SSP). The excellent match between the observed and the
synthetic spectrum implies that the bulk of stars in the four galaxies
had to form at $z\gsim 2.5-3$, given that they lie at $1.6<z<1.9$.
Indeed, the power of the Mg-UV feature is that it allows us to get at
once the redshift {\it and} an age estimate, hence to set a formation
redshift.  This is illustrated in Figure 3, showing that indeed the
feature appears only in populations that have evolved passively since
at least $\sim 0.5$ Gyr, while the feature is absent for younger ages
or for populations with ongoing star formation. Therefore, this
feature being observable up to high redshifts is very useful for
age-dating PEGs (see e.g., Maraston et al. 2006).

\begin{figure}
\centering
\includegraphics[height=8cm]{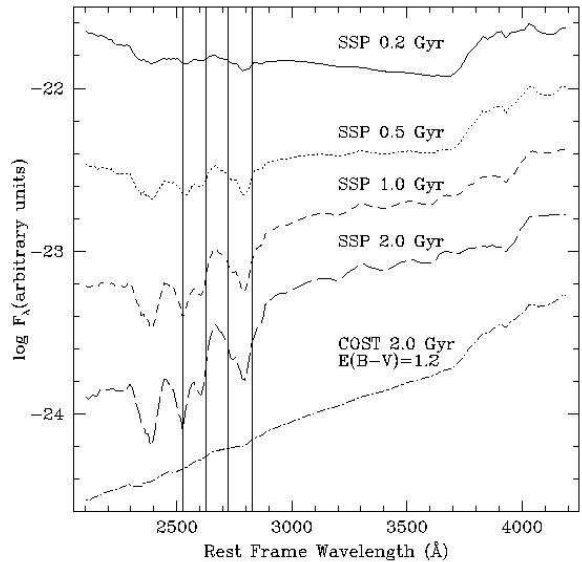}
%\picplace{5cm}{2cm} % Give the correct figure height and width in cm
%
\caption{The synthetic UV spectrum of a simple stellar population at
several ages as indicated (from Daddi et al. 2005). Note that the
relative fluxes at the various epochs correspond to the passive fading
of the stellar population.  The spectrum for the case of a
time-constant star formation rate is also shown. (Synthetic models
are from the library of Bruzual \& Charlot 2003.)}
\label{fig:3}       % Give a unique label
\end{figure}

Thanks to the Mg-UV feature soon more high-$z$ PEGs were discovered
(McCarthy et al.  2004; Daddi et al. 2005), but still within
relatively small fields and capitalizing on spectroscopic surveys not
specifically targeting such objects. For a more focused spectroscopic
study of high-$z$ PEGs an effective criterion for selecting candidates
was clearly required.

\section{The $BzK$ Criterion for Selecting High-$z$ PEGs}

An efficient photometric criterion for the selection of PEGs at high
redshift was proposed by Daddi et al. (2004), based on the expected
location of $z>1.4$ PEGs on the $(z-K)$ vs $(B-z)$ two-color
diagram. Moreover, the criterion was validated on existing
spectroscopic data (K20, GOODS, etc.). The $BzK$ plot from Kong et
al. (2006) is shown in Figure 4. Kong et al. then estimated that the
comoving volume density of ``passive BzKs'' (pBzKs) with
$M_*>10^{11}\msun$ and $<\!z\!>\simeq 1.7$ is only $\sim 20\pm7\%$ of the
corresponding space density at $z=0$. Since the majority of such
massive PEGs appears to be already in place at $z\sim 1$, this implies
that the epoch between $z\sim 2$ and $z\sim 1$ ($\sim 2.5$ Gyr) must
be that during which most of massive PEGs are either finally
assembled, and/or any residual star formation in their precursors is
shut off.

\begin{figure}
\centering
\includegraphics[width=10cm, angle=-90]{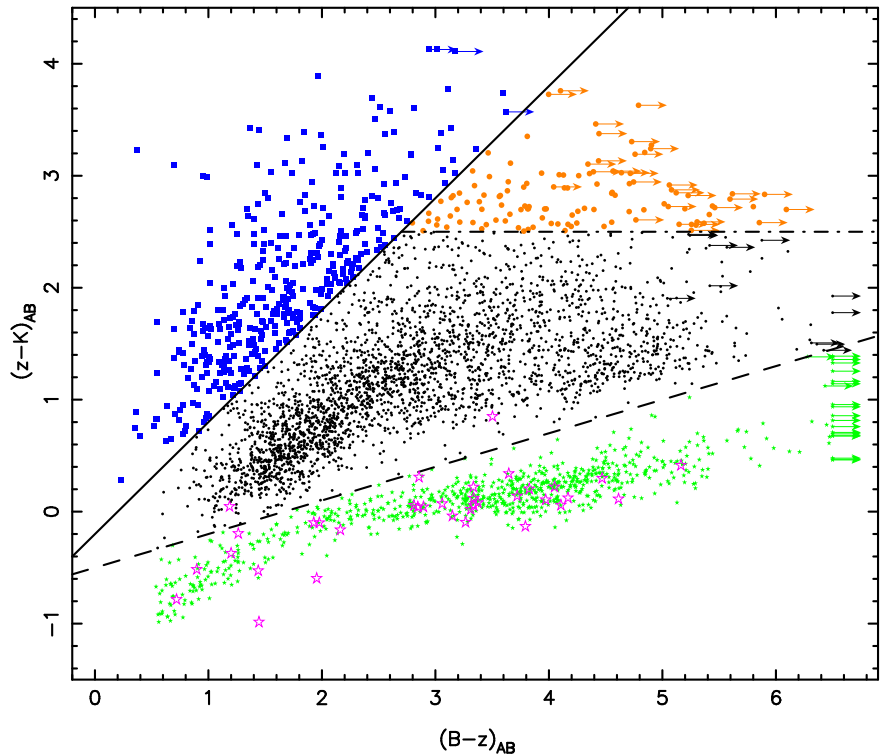}
%\includegraphics[height=6cm]{cullingpbzks.ps}
%\picplace{5cm}{2cm} % Give the correct figure height and width in cm
%
\caption{The BzK two-color plot from Kong et al. (2006). Following
Daddi et al. (2004), four distinct regions are identified, each
dominated by PEGs at $z>1.4$, or pBzKs (upper right, red symbols),
star-forming galaxies at $1.4<z<2.5$ or sBzKs (upper left, blue
symbols), galaxies at $z<1.4$ (middle, black symbols), and stars
(lower section, green symbols). Purple stars indicate Galactic stellar
templates.}
\label{fig:4}       % Give a unique label
\end{figure}

Besides high-$z$ PEGs, the BzK criterion is also very efficient at
selecting actively star-forming galaxies at $1.4<z<2.5$ (hence called
sBzKs), even if rather heavily reddened. The early zCOSMOS
spectroscopic results have demonstarted that indeed the contamination
by lower redshift interlopers is extremely small for sBzK-selected
targets (Lilly et al. 2007). Thus, a sizable fraction of sBzKs must
turn into pBzKs and lower redshift PEGs between $z\sim 2.5$ and $z\sim
1$. How this transformation is taking place, in terms of
star-formation quenching, merging, disk instability, and morphological
transformation is perhaps the major open issue concerning the
evolution of galaxies over the last $\sim 11$ Gyr of cosmic time. Of
course, this must include the co-evolution of hosting galaxies and
their central supermassive black holes (SMBH), and a first grasp at
the AGN activity among sBzK galaxies and the concomitant growth of
their stellar mass and SMBH can be found in Daddi et al. (2007).

\begin{figure}
\centering
\includegraphics[height=7.5cm]{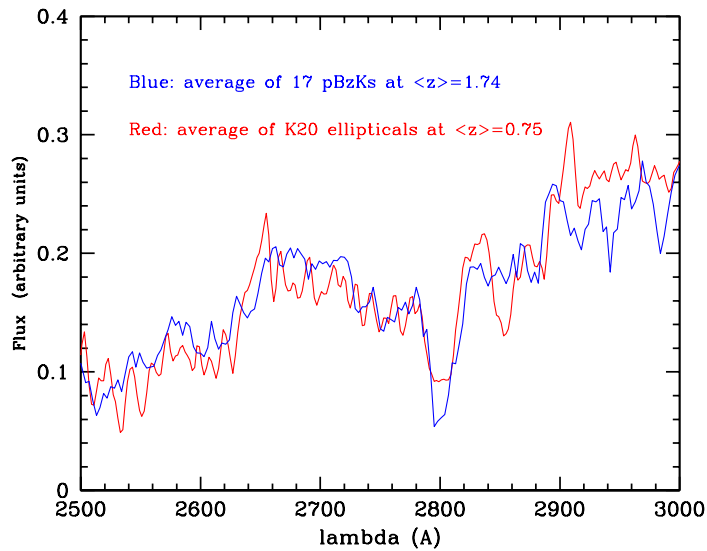}
%\picplace{5cm}{2cm} % Give the correct figure height and width in cm
%
\caption{The coadded rest-frame UV spectrum for 17 pBzKs from the
sample of Kong et al. (2006) obtained at the VLT with the VIMOS
spectrograph (Daddi et al.  in preparation). Also shown is the coadded
spectrum of K20 PEGs at $z\simeq 0.8$, reproduced from Fig. 1.}
\label{fig:5}       % Give a unique label
\end{figure}

\section{More Rest-frame UV Spectroscopy of High-$z$ PEGs}

While extensive multi-object spectroscopy of star-forming galaxies at
$z\gsim 1.4$ is relatively easy nowadays (e.g. with VIMOS and FORS2 at
the VLT and LRIS at Keck), an extensive spectroscopic survey of
high-$z$ PEGs is still lacking. I like to report here some preliminary
result from a couple of {\it pilot} projects, aiming primarily at
understanding how to better plan for future massive surveys.

\begin{figure}
\centering
\includegraphics[height=7.5cm]{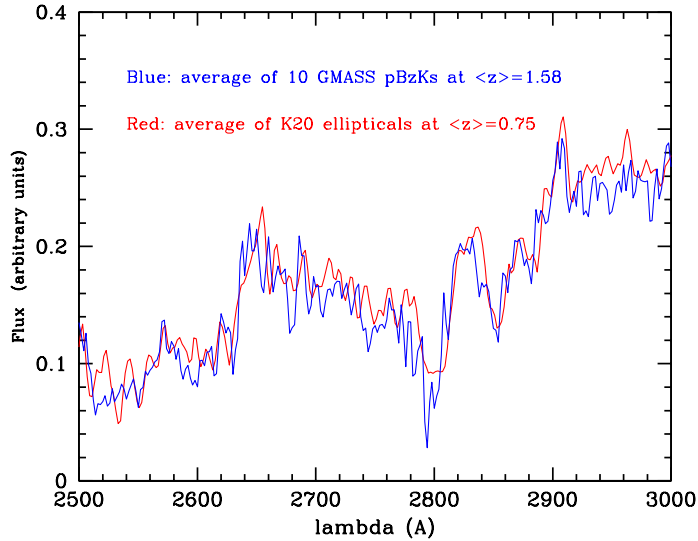}
%\picplace{5cm}{2cm} % Give the correct figure height and width in cm
%
\caption{The coadded rest-frame UV spectrum for 10 pBzKs from the
GMASS project obtained at the VLT with the FORS2
spectrograph (Cimatti et al.  in preparation). Also shown is the coadded
spectrum of the K20 PEGs reproduced from Fig. 1.}
\label{fig:6}       % Give a unique label
\end{figure}

Figure 5 shows the coadded spectrum for 17 pBzKs selected from the
sample of Kong et al. (2006), and observed with the VIMOS spectrograph
at the VLT with an integration time of 2.5h (Daddi et al. in
preparation). The spectra were taken with the intermediate resolution
($R\sim 600$) Red Grism, while targeting sBzKs and other galaxies as
well, so to fully exploit the multiplex of the instrument. Overplotted to
the coadded spectrum of the pBzKs is the template spectrum for the
$z\simeq 0.8$ PEGs from the K20 survey (from Mignoli et al. 2005).

Figure 6 shows the coadded spectrum for 10 pBzKs selected from the
GMASS sample, and observed with the FORS2 spectrograph at the VLT with
an integration time of 30h (sic!)  (Cimatti et al. in preparation; Kurk et
al. in preparation). The spectra were taken with a low resolution ($R\sim
300$) Red Grism. The same lower redshift PEG spectrum is also
overplotted. Note that the coadded spectrum in this figure has the same S/N
an individual galaxy would have if integrated for 300 hours.

Individual spectra taken with FORS2 have far better S/N than those
taken with VIMOS, given the much longer integration time, and the
superior performance of FORS2 for $\lambda\gsim 8000$ \AA. In practice, 
we are fairly secure that all 10 FORS2 redshifts are correct, while a few
(perhaps 3 or 4, or $\sim 20\%$) among the VIMOS redshifts may not be correct.

\section{Conclusions}

Optical spectroscopic observations in the rest-frame UV have
demonstrated that PEGs exist in sizable number up to at least $z\sim
2$. The surface density of $BzK$-selected PEGs is $\sim 0.2$
arcmin$^{-2}$ down to $K=20$, substantially lower than the
slit-packing capability of multiobject spectrographs such as VIMOS or
FORS2.  Moreover, the rest-frame UV continuum is quite naturally very
faint in PEGs, even in high-redshift ones, given that for being
passive these galaxies lack OB stars in sizable number. Actually, the
absence of these stars is a requisite for detecting the Mg-UV feature.

Therefore, none of the optical instruments currently at 8-10m
telescopes is optimally suited to exploit the Mg-UV feature for
mapping the population of high-$z$ PEGs. Very long integration times
would be required, coupled with only a marginal exploitation for the
specific targets of the instrument multiplex.

An alternative option to optical spectroscopy is to go to the near-IR,
following the CaII doublet and the 4000 \AA\ break as they redshift
beyond $\sim 1\mu$.  However, the secure detection even of the
strongest absorption lines such as CaII H\&K would need extremely long
integration times, which could be afforded only if very many objects
could be observed simultaneously.  The only near-IR MOS currently in
operation is MOIRCS at the SUBARU telescope (Tokoku et
al. 2006). Several BzK-selected galaxies from the sample of Kong et
al. (2006) were ready to be observed with MOIRCS during a 5-night run
in April 2007. Unfortunately the spectrographic channel of MOIRCS had
a technical problem when just inserting the first cryogenic mask into
the instrument, and only the imaging mode could be used in the
following nights.  As well know, cryogenics is critical for operations
in the $K$ band, but a warm instrument working just in the $J$ band
would be perfectly adequate for observing the CaII doublet and the
4000 \AA\ break up to $z\sim 2$ and beyond. A first opportunity in
this sense will possibly be offered once more by the SUBARU telescope,
using the FMOS multifiber instrument (Kimura et al.  2003). With its
operation over the large FoV at the prime focus of SUBARU, FMOS will
have a very good match between its multiplex and the surface density
of pBzKs, and therefore promises to be perhaps a better option for
mapping the population of high-$z$ PEGs before JWST. Thus,
spectroscopy in the rest-frame ultraviolet has been instrumental for
demonstrating the existence of passively evolving galaxies at high
redshifts, as well as for validating an efficient photometric
criterion for selecting them. However, it appears that near-IR
multiobject spectroscopy targeting classical features in the
rest-frame optical such as the CaII doublet and the 4000 \AA\ break
may be better suited for a massive survey of these objects.

\medskip

I would like to thank my colleagues Andrea Cimatti, Emanuele Daddi,
and Marco Mignoli for their permission to show the coadded spectra
shown in Figure 5 and 6 in advance of publication.

%
%

%%%%%%%%%%%%%%%%%%%%%%%%%%%%%%%%%%%%%%%%%%%%%%%%%%%%%%%%%%%%%%%%%%%%%%  }

%\printindex
\end{document}